\documentclass[aps,pre,twocolumn]{revtex4-1}
\usepackage{graphicx}
\usepackage{dcolumn}
\usepackage{amsmath}
\usepackage{bm}
\usepackage{hyperref}
\usepackage{latexsym}
\usepackage{verbatim}
\usepackage{color}
\usepackage{subfigure}
\usepackage{bm}

\begin{document}

\title{Self-regulation in Self-Propelled Nematic Fluids}
\author{Aparna Baskaran$^{1}$ and M. Cristina Marchetti$^{2}$}
\affiliation{$^{1}$Martin Fisher School of Physics, Brandeis University, Waltham, MA, USA.\\
$^{2}$Physics Department and Syracuse Biomaterials Institute, Syracuse University, Syracuse, NY 13244, USA.\\}
\date{\today }

\begin{abstract}
We consider the hydrodynamic theory of an active fluid of self-propelled particles
with nematic aligning interactions. This class of materials has polar symmetry at the microscopic level, but forms macrostates of nematic symmetry. We highlight three key features of the dynamics.   First, as in polar active fluids, the control parameter for the order-disorder transition, namely the density, is dynamically convected by active currents, resulting in a generic, model independent dynamical self-regulation that destabilizes the uniform nematic state near the mean-field transition.
Secondly, curvature driven currents render the system unstable deep in the nematic state, as found previously.  Finally, and unique to self-propelled nematics, nematic order induces local polar order that in turn leads to the growth of density fluctuations.  We propose this as a possible mechanism for the smectic order of polar clusters seen in numerical simulations.
\end{abstract}

\maketitle

\section{Introduction}
\label{sec:intro}

Active materials are soft materials driven out of equilibrium by energy
input at the microscale. This liberates the fluctuations from the
constraints of equilibrium such as fluctuation-dissipation relations and
reciprocity. As a consequence, several exotic emergent behaviors result,
such as long range order in 2D \cite{Toner1995, Toner1998}, anomalous
fluctuations \cite{Ramaswamy2003,Narayan2007}, dynamical structures and patterns \cite
{Koch1994,Budrene1991,Budrene1995,Matsushita1997}.
In addition to serving as prototypical systems to explore emergent dynamical behavior,
active materials also form the physical scaffold of
biological systems in that active matter, when coupled to regulatory
signaling pathways, provides a model for a variety of living systems,
such as bacterial biofilms or the cytoskeleton of a cell.

\begin{figure}[tbp]
\includegraphics [width=8cm] {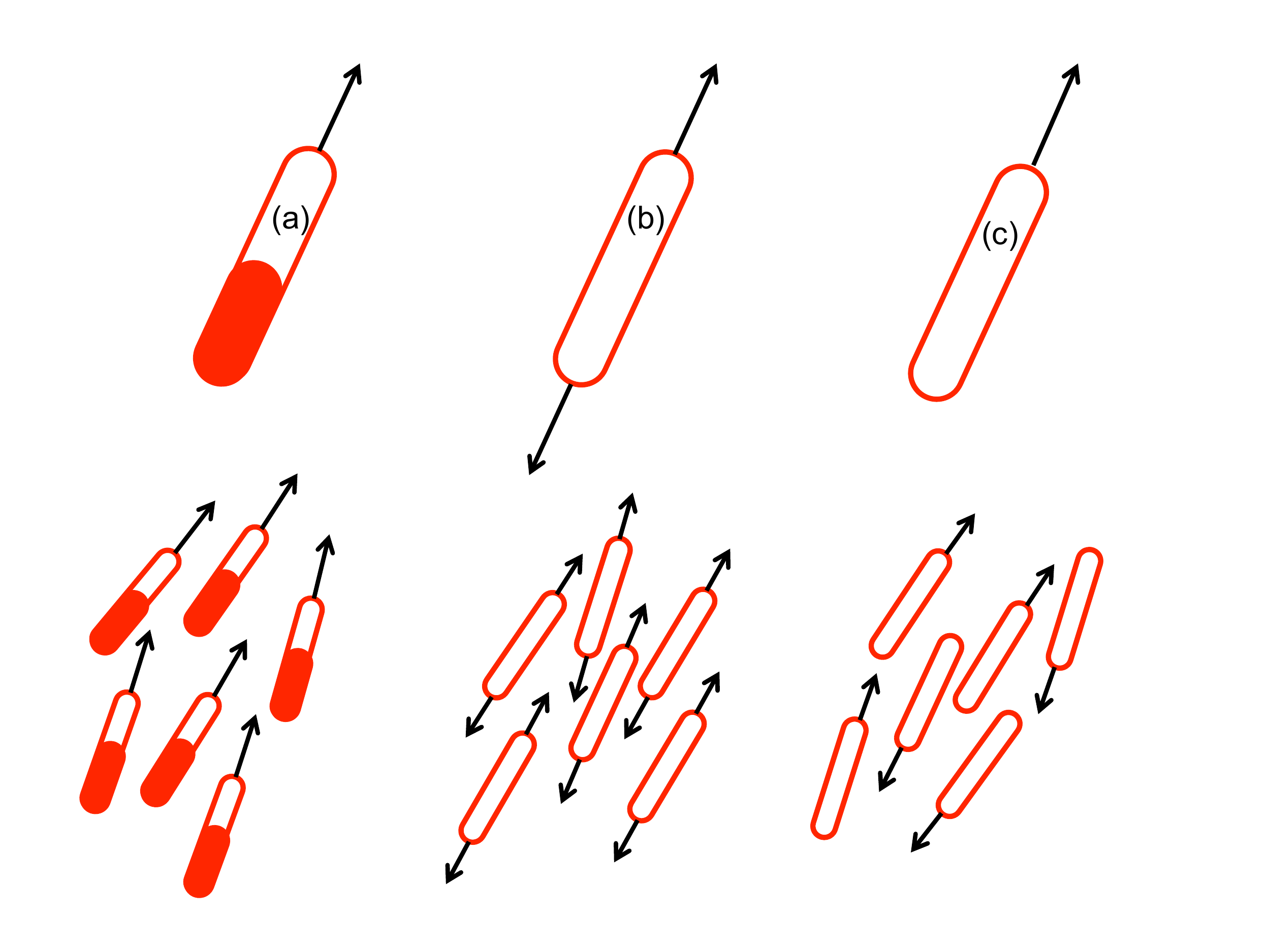}
\caption{ Top - active particles of various microscopic symmetry: (a) Polar active particles with head/tail asymmetry resulting in polar interactions, as studied in \cite{Vicsek1995,Gregoire2004}, (b) Apolar active particles, as studied in \cite{Simha2002,Ramaswamy2003}, (c) Sels-propelled particles, with physical head-tail symmetry, resulting in apolar interactions. Bottom - ordered macroscopic states of active particles: polar active fluid (left) formed by polar particles (a) with polar interactions; active nematic fluid (center) formed by apolar particles (b) with apolar interactions; self-propelled active nematic fluid (right) formed by self-propelled particles (c) with apolar interactions.}
\label{Fig1}
\end{figure}

Active particles are generally elongated and form orientationally ordered states~\cite{Ramaswamy2010}. The nature of the  ordered state depends on both the symmetry of the individual
particles and the symmetry of the aligning interactions (see Fig.~\ref{Fig1}). Physical realizations of \emph{polar} active  particles (characterized by a head and a tail)  include bacteria, asymmetric vibrated granular rods, and polarized migrating cells. Polar active entities are often modeled as self-propelled (SP) particles,
where the activity is incorporated via a self-propulsion velocity of
the individual entities. \emph{Apolar} (head-tail symmetric)  active particles, often referred to as "shakers", have also been considered in the literature. Realizations are symmetric vibrated rods~\cite{Narayan2007}. It has also been argued that melanocytes, the cell that distribute pigments in our skins, may effectively behave as "shakers"~\cite{Gruler1999,Kemkemer2000}.

The nature of the interaction is of course crucial in controlling the properties of the ordered state.
Apolar active particles generally experience apolar interactions and the resulting ordered state has the symmetry of equilibrium nematic liquid crystals. The broken orientational symmetry identifies a direction ${\bf \hat{n}}$, but the ordered state is invariant for ${\bf \hat{n}}\rightarrow -{\bf \hat{n}}$. The properties of these \emph{active nematic} fluids  have been studied by several authors~\cite{Simha2002,Ramaswamy2003}.
SP  and polar particles  may experience either polar interactions, i.e., ones that tend to align particles head to head and tail to tail, or interactions that are apolar, i.e, align particles regardless of their polarity. A well studied example of polar particles with polar interactions is provided by Vicsek-type models~\cite{Vicsek1995,Gregoire2004}.
This class of active systems can order in polar states, characterized by a nonzero vector order parameter and mean motion, and will be referred to as \emph{active polar}  fluids.

One can envisage a third class of active fluids that consists of SP particles whose emergent macrostates are nematic in symmetry. A realization of this is self-propelled particles with physical interactions, such as steric repulsion or hydrodynamic interactions among swimmers in a suspension. It has been shown that these interactions lead to nematic, rather than polar order~\cite{Baskaran2008a,Baskaran2009}. What is key is the fact that interactions such as steric collisions or hydrodynamic couplings individually conserve momentum
and hence cannot lead to the development of a macroscopic momentum for the system.  The ordered state of these systems has nematic symmetry, but as we will see below its properties are distinct from those of active nematics composed of apolar particles. Here we refer to this third class of active systems as \emph{self-propelled nematic} fluids. Further, it has recently become apparent that models of polar particles with apolar interactions may be  relevant  to a number of physical systems, including gliding myxobacteria~\cite{Peruani2012},
suspensions of self-catalythic Janus colloids~\cite{Palacci2010} and motile epithelial cell sheets such as those studied in wound healing assays~\cite{Saez2007,Petitjean2010}. Self-propelled nematics therefore represent  an important new class of active systems
of direct experimental relevance.

A useful theoretical framework for describing the collective behavior of active systems
is a  continuum model that generalizes liquid crystal hydrodynamics to include new terms induced by activity~ \cite{Simha2002,Kruse2004,Voituriez2005,Voituriez2006,Marenduzzo2007,Giomi2008,Cates2008,Sokolov2009,Giomi2010,Saintillan2010}.  In this paper, we consider a minimal
phenomenological  continuum model of self-propelled nematics based on the equations derived by us from a microscopic model of SP hard rods. We show that in these systems it is important to explicitly retain the dynamics of both the  collective velocity or polarization field of the particles and of the nematic order parameter to
unfold the mechanisms at play in the formation of emergent structures. Further, we show that \emph{all} active nematics (both consisting of shakers and SP particles) exhibit the phenomenon of dynamical self regulation, due to the fact that the parameter controlling the order-disorder transition, namely the density   $\rho $
of active particles, is not externally tuned, as in systems undergoing equilibrium phase transitions, but it is
dynamically coupled to the order
parameter. This coupling is analogue to the one present in polar fluids~\cite{Toner1995, Toner1998,Toner2005} and is a generic mechanism for emergent structure in all active systems, as demonstrated in our recent work~\cite{Gopinath2012}.

The layout of the paper is as follows. First, we construct the hydrodynamic
description of a self-propelled  nematic using symmetry
considerations highlighting the key features that distinguish the dynamics of the system from that of an active nematic. Then we examine the linear stability of  the homogeneous nematic state. We show that there exists three dynamical mechanisms responsible for emergent structures in active fluids with nematic symmetry. The first is a
model-independent instability that occurs in the vicinity of the mean field order
disorder transition due to the coupling between order parameter and mass transport
which renders the dynamics of the system self regulating. We argue that this instability
is the basis for the emergence of bands and phase separation
found ubiquitously active systems ~\cite{Chate2006,Yang2010,Ginelli2010}.  The second is the well known instability of director fluctuations that arises from  nonequilibrium curvature-induced fluxes and is closely related to the giant number fluctuations observed in these systems~\cite{Simha2002,Ramaswamy2003}.  These two instabilities are common to both active nematics and self-propelled nematics, i.e., occur regardless of the symmetry of the microdynamics.  Finally, we show that there exists a third instability unique to self-propelled nematic fluids due to the fact that
in these systems, large scale nematic order can induce local  polar order, which  in
turn destabilizes the density. This mechanism may be responsible for the smectic order of polar clusters observed recently in simulations of SP rods~\cite{Yang2010,Sam2012}. We conclude with a brief discussion.

\section{The Macroscopic Theory}
\label{sec:macro}

The hydrodynamic equations of a self-propelled nematic
have been derived
from systematic coarse-graining of specific microscopic models~\cite{Baskaran2008a,Baskaran2009}.
Here we  introduce these equations phenomenologically, with the goal of examining the dynamics
without the limitations imposed by the specific parameter values obtained from a microscopic model or resulting from the choice of the closure used in the kinetic equation.

We limit ourselves to overdamped systems in two dimensions.
The hydrodynamic equations are then written in terms of
 three continuum fields: the conserved number density  $\rho \left( \mathbf{r},t\right) $ of
active units, the polarization density $\bm\tau({\bf r},t)=\rho({\bf r},t)
\mathbf{P}\left( \mathbf{r},t\right) $, with $\mathbf{P}\left( \mathbf{r},t\right)$ a polarization order parameter, and
the nematic alignment density tensor $Q_{ij}({\bf r},t)=\rho({\bf r},t)S_{ij}\left( \mathbf{r},t\right) $. The polarization ${\bf P}$ is directly proportional to   the collective velocity of the active particles, while   $S_{ij}$ is the conventional  nematic order
parameter tensor familiar from liquid crystal physics. For a uniaxial system in two dimensions, $Q_{ij}$ is a symmetric traceless tensor with only two independent components
and can be written in terms of a unit vector ${\bf \hat{n}}$ as $Q_{ij}=Q(\hat{n}_i\hat{n}_j-\frac12\delta_{ij})$, where $Q=\rho S$; $S$ is  the magnitude of the order parameter and the director  ${\bf \hat{n}}$ identifies the direction of spontaneously broken symmetry in the nematic state.    For simplicity most of the discussion below refers to the case where the active particles are modeled as long thin rods with repulsive interactions.

\subsection{Active Nematic Hydrodynamics}
\label{subsec:active_nematic}
We first  construct the dynamical equations of the system including single-particle convection terms induced by self propulsion, but assuming that self-propulsion does not modify the interaction between two rods. The  hydrodynamic equations then take the form~\cite{Baskaran2008}
\begin{subequations}
\begin{gather}
\partial _{t}\rho +v_{0}\nabla \cdot \bm{\tau }=D\nabla ^{2}\rho
\label{1.1}\\
\partial _{t}\bm{\tau }+D_r\bm{\tau }+v_{0}\nabla \cdot \mathbf{Q}+
\frac{v_{0}}{2}\nabla \rho =D_\tau\nabla ^{2}\bm{\tau }  \label{1.2}
\end{gather}
\begin{align}
\partial _{t}Q_{ij}-&D_r\left[ \alpha\left( \rho \right) -\beta
{\bf Q}{\bf :}{\bf Q}\right] Q_{ij}+v_{0}F_{ij} = D_{b}\nabla ^{2}Q_{ij}\notag\\
&+D_{s}\partial _{k}\left( \partial _{i}Q_{kj}+\partial
_{j}Q_{ik}-\delta _{ij}\partial _{l}Q_{kl}\right)\notag\\
 &\label{1.3}
 \end{align}
\end{subequations}
where ${\bf Q}{\bf :}{\bf Q}=Q_{kl}Q_{kl}$, $D_r$ is the rotational diffusion rate,  and
$F_{ij}=\left( \partial _{i}\tau _{j}+\partial _{j}\tau _{i}-\delta
_{ij}\nabla \cdot \tau \right)$. All terms proportional to $v_0$ arise from one-particle convection due to self-propulsion and are the only consequence of activity in this simple model. The repulsive
interactions among the particles  generate the cubic homogeneous term (with $\beta>0$) on the right hand side of Eq.~\eqref{1.3} and a change in sign of $\alpha(\rho)\sim \rho-\rho_c$ at a critical density $\rho_c$, controlling the transition between the isotropic and the nematic states. ~\footnote{Note that the cubic term was not derived in Ref.~\cite{Baskaran2008}, but is easily obtained by a higher order closure of the moment expansion of the kinetic equation.}
Interactions also give density-dependent corrections to the various diffusion coefficients for density ($D$), polarization ($D_\tau$),
splay ($D_s$) and bend ($D_b$) deformations of the nematic alignment tensor.
We will
ignore all such corrections in the following~\footnote{{ Retaining the density dependence of the diffusion coefficients results in interesting emergent structures as shown by \cite{Cates2010}. Since we seek to focus on fundamental features that do not depend on the detailed structure of the hydrodynamic coefficients, we ignore this physically important feature.}}. Due to the fact
that the interactions are purely nematic, the polarization $\bm \tau $ decays on short time scales $\sim D_r^{-1}$
for all strengths of activity. At long time, a hydrodynamic description can then be obtained by neglecting $\partial_t\bm\tau$ in Eq.~\eqref{1.2}, and using Eq.~\eqref{1.2} to eliminate $\bm\tau$ from the other equations, with the result (to leading order in gradients)
\begin{subequations}
\begin{gather}
\partial _{t}\rho =D\nabla ^{2}\rho+{\cal D}_Q\bm\nabla\bm\nabla{\bf :}{\bf Q}
\label{2.1}
\end{gather}
\begin{align}
\partial _{t}Q_{ij}-&D_r\left[ \alpha\left( \rho \right) -\beta
{\bf Q}{\bf :}{\bf Q}\right] Q_{ij} = D_{b}\nabla ^{2}Q_{ij}\notag\\
&+D_{s}\partial _{k}\left( \partial _{i}Q_{kj}+\partial
_{j}Q_{ik}-\delta _{ij}\partial _{l}Q_{kl}\right)\notag\\
&+{\cal D}_\rho( \partial _{i}\partial _{j}-\frac12\delta _{ij}\nabla
^{2}) \rho
 &\label{2.2}
 \end{align}
 \label{2}
\end{subequations}
This procedure yields qualitative new terms proportional to ${\cal D}_Q$ and ${\cal D}_\rho$ that vanish in equilibrium.
The term proportional to ${\cal D}_Q$ is the curvature induced density flux that has been discussed extensively by Ramaswamy and collaborators~\cite{Ramaswamy2003} and shown to be
responsible for
giant number fluctuations in the ordered state of active nematic. The diffusive coupling proportional to ${\cal D}_\rho$ describes similar physics but has not been considered in earlier description of active nematic fluids.  In addition, activity also yields corrections to the various diffusion coefficients.
We have, however, implicitly neglected those here by retaining the same notation for these quantities as in Eqs.~\eqref{1.1}-\eqref{1.3} to highlight the difference between these corrections that do not change the dynamics qualitatively and the new terms
proportional to ${\cal D}_Q$ and ${\cal D}_\rho$.  Although obtained here by considering a system of self-propelled particles, Eqs.~\eqref{2.1} and \eqref{2.2} have the same structure as the hydrodynamic equations of  an active nematic, consisting of a collection of \emph{apolar} active particles (shakers) with apolar interactions. This is an important point as it stresses that the qualitative differences between active and self-propelled nematic that have been observed in simulations must arise entirely from the dependence of the interaction on self propulsion $v_0$.

\subsection{Momentum-conserving interaction of self-propelled nematogens}
\label{subsec:interaction}
As shown in Ref.~\cite{Baskaran2008a} and supported by simulations of self-propelled hard rods~\cite{Peruani2006,Ginelli2010,Yang2010},  self-propulsion does modify the repulsive interaction in a qualitative way. This modification results in  local build-up of polarization in the nematic state, making it necessary to retain the dynamics of polarization density  in the continuum model.
\begin{figure}[tbp]
\includegraphics [width=0.49\textwidth] {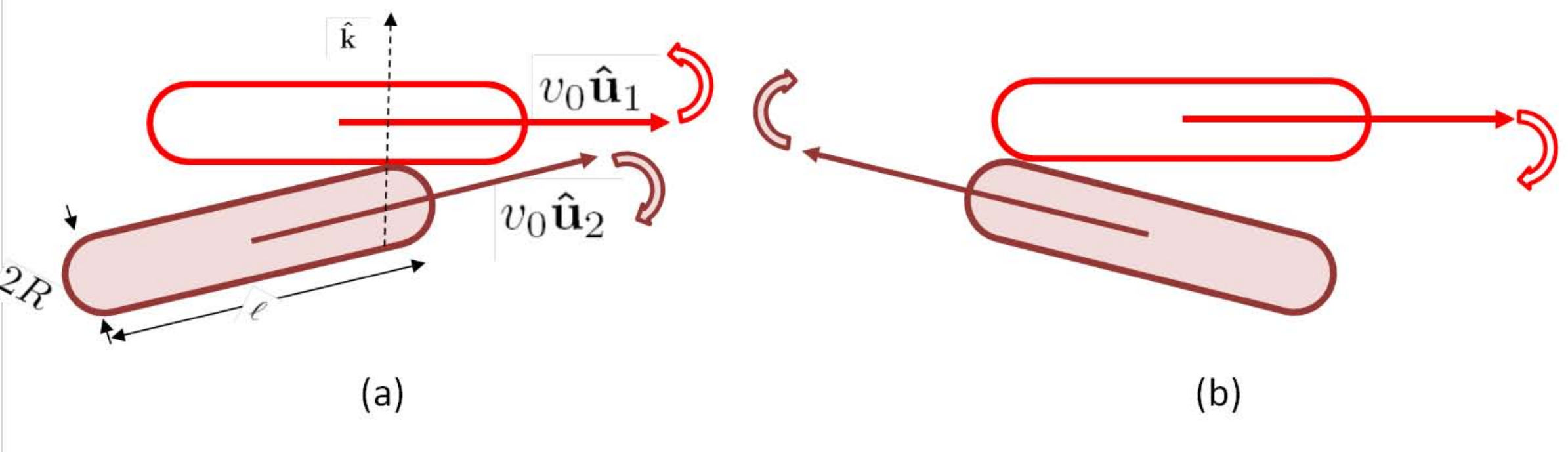}
\caption{ Illustration of momentum conserving collisions among self propelled
particles. It can readily be shown that two rods as shown in (a),
coming in with only their self-replenishing velocities, will acquire opposite
angular momenta $\bm\omega _{1}\sim {\bf \hat{z}} \ell v_{0}\left[{\bf \hat{z}}\cdot\left( \mathbf{\hat{u}}\right]%
_{1}\times \mathbf{\hat{u}}_{2}\right) $ and $\bm\omega _{2}\sim-{\bf \hat{z}} \ell
v_{0} \left(
\mathbf{\hat{u}}_{1}\cdot \mathbf{\hat{u}}_{2}\right)\left[{\bf \hat{z}}\cdot\left( \mathbf{\hat{u}}_{1}\times \mathbf{\hat{u}}_{2}\right)\right] $, where the vectors
are defined in the image and in ~\cite{Baskaran2010}. The collision will therefore induce rotations as indicated, promoting alignment of the two rods.
On the other hand, two nearly antialigned rods as in (b) acquire angular momenta of the same sign, inducing rotation of both rods in the same directions,
and leaving their relative angle unchanged. }
\label{Fig1.1}
\end{figure}
The modification of the Onsager excluded volume interactions among hard rods due to self propulsion is worked out  in Ref.~\cite{Baskaran2010}.
Here we simply give a qualitative description of this effect  and we refer the reader to that work for the technical details.
First we note that the presence of a self propulsion speed along the long axis of
the nematogen, results in a breaking of the nematic symmetry of the the microdynamics, as shown in Fig.~\ref{Fig1}.
On the other hand, since the
interactions conserve momentum, this cannot lead to a macroscopic breaking
of polar symmetry as this would amount to the appearance of a spontaneous
macroscopic momentum from a zero momentum state. Hence, only a  homogeneous  ordered nematic state
can occur and the associated mean field transition will be the same as in the
case of the active nematic considered above, albeit with coefficients $
\alpha$ and $\beta$ renormalized by self-propulsion~\cite{Baskaran2008a}.
Even though the polar symmetry cannot be broken macroscopically,
momentum conservation allows the nematic ordering to induce local polar order
in the system. To illustrate this, let us consider hard rods in two dimensions
undergoing energy-momentum conserving interactions. As shown in Fig. (\ref
{Fig1.1}), the angular momentum transfer due to the linear
momentum from self-propulsion for a collision between two rods scales as $\omega \sim \cos \left(
\theta _{1}-\theta _{2}\right) \sin \left( \theta _{1}-\theta _{2}\right) $.
If the rods are nearly aligned head to head (as in Fig \ref{Fig1.1}.a), the effect of
this angular momentum is to turn the rods towards each other, while if they are
nearly aligned head to tail as in Fig \ref{Fig1.1}.b, the collision turns both rods in the same direction, leaving their relative angle unchanged. This mechanism effectively promotes head-to-head alignment. Since collisions among such
nearly aligned nematogens will  dominate the dynamics  in the nematic
state, the nematic order effectively induces polar order.

\subsection{Self-propelled Nematic Hydrodynamics}
\label{subsec:sp_nematic}
The fact that interactions among self-propelled nematogens tend to induce polar order is reflected in the hydrodynamic description by a number
of new nonlinear terms that couple $\bm\tau$ and ${\bf Q}$, with coefficients that vanish in the limit $v_0=0$.
The continuum
equations for a self-propelled nematic that incorporate the above physics are given by
\begin{widetext}
\begin{subequations}
\begin{gather}
\label{3.1}
\partial _{t}\rho +v_{0}\nabla \cdot \bm{\tau }=D
\nabla ^{2}\rho +\mathcal{D}_{Q}\nabla \nabla :\mathbf{Q}\\
\partial _{t}\bm{\tau }+D_r\bm{\tau }+\gamma_1{\bf Q}{\bf :}{\bf Q}\bm\tau-\gamma_2\bm{\tau \cdot Q}
+\lambda_1 \bm{\tau }\cdot \nabla \bm{\tau } \mathbf{=}-v_{0}\bm\nabla
\cdot \mathbf{Q}-\frac{v_{0}}{2}\nabla \rho
+\lambda _{2}\bm{\tau }\nabla \cdot \bm{\tau }+\frac{\lambda _{3}}{2}\nabla
\tau ^{2}+D_{\tau}\nabla ^{2}\bm{\tau }
\label{3.2}\\
\partial _{t}Q_{ij}-D_r(\alpha-\beta {\bf Q}{\bf :}{\bf Q}) Q_{ij}+v_0F_{ij}+\lambda _{4}G_{ij} =
\mathcal{D}_\rho( \partial _{i}\partial _{j}-\frac{\delta _{ij}}{2}%
\nabla ^{2}) \rho
+D_{s}\partial _{k}\left( \partial _{i}Q_{kj}+\partial
_{j}Q_{ik}-\delta _{ij}\partial _{l}Q_{kl}\right)
+D_{b}\nabla ^{2}Q_{ij}
\label{3.3}
\end{gather}
\label{3}
\end{subequations}
\end{widetext}
where again we have implicitly neglected active corrections to $D$, $D_\tau$, $D_s$ and $D_b$ to highlight the new, purely active terms.
Activity enters in Eqs.~\eqref{3}  through the convective terms proportional to $
v_{0}$, the new terms with coefficients  $\gamma_i$ and to $\lambda_i $,
which vanish in equilibrium, as well as the terms proportional to ${\cal D}_Q$ and ${\cal D}_\rho$ that arise here from active corrections to interactions. Finally,  the parameters
 $\alpha$
and $\beta$  controlling the mean field isotropic-nematic transition are also renormalized by activity.
The homogeneous nonlinearities proportional to $\gamma_i$  in the
polarization equation encode the fact that nematic order induces
polar order. The latter is, however, only local as the equations do not a admit a homogeneous solution with nonzero $\bm\tau$. Further, the active modification of the interactions, yield the convective nonlinearities $\sim{\cal O}(\bm{\tau }\nabla \bm{\tau })$ that play a central role in the emergent physics of active polar fluids.

Since the goal of this presentation is to highlight the mechanisms
responsible for emergent structures, we simplify the equations by setting all of the equilibrium-like
diffusion coefficients to be equal, i.e., $D=D_\tau=D_b=D_{0}$, with the exception of the splay relaxation constants $D_{s}$.
 In addition, we assume $\lambda_i=\lambda $ for all $i$'s and
$\gamma_1=\gamma_2=\gamma$.  Finally, we measure time in units of $1/D_r$ and lengths in units of $
\sqrt{D_{0}/D_r}$. The hydrodynamic equations then simplify to (in
nondimensional form)
\begin{widetext}
\begin{subequations}
\begin{gather}
\partial _{t}\rho +\overline{v}\bm\nabla \cdot \bm\tau=\nabla ^{2}\rho +
\overline{D}_Q\bm\nabla \bm\nabla :\mathbf{Q}  \label{4.1}\\
\partial _{t}\bm{\tau }+\left(1+\gamma{\bf Q}{\bf :}{\bf Q}\right)\bm\tau-\gamma\bm{\tau \cdot Q}
+\lambda \bm{\tau }\cdot \nabla \bm{\tau } =-\overline{v}
\bm\nabla \cdot \mathbf{Q}-\frac{\overline{v}}{2}\bm\nabla \rho
+\lambda \left( \bm{\tau }\nabla \cdot \bm{\tau }+\frac12\nabla \tau
^{2}\right) +\nabla ^{2}\bm{\tau }  \label{4.2} \\
\partial _{t}Q_{ij}-\left( \alpha-\beta{\bf Q}{\bf :}{\bf Q}  \right) Q_{ij}+\overline{v} F_{ij}+\lambda G_{ij} =
\overline{D}_\rho( \partial _{i}\partial _{j}-\frac12 \delta _{ij}\nabla ^{2}) \rho
+\overline{D}_{S}\partial _{k}\left( \partial _{i}Q_{kj}+\partial
_{j}Q_{ik}-\delta _{ij}\partial _{l}Q_{kl}\right)
+\nabla ^{2}Q_{ij}
\label{4.3}
\end{gather}
\label{4}
\end{subequations}
\end{widetext}
with $\overline{D}_Q={\cal D}_Q/D_0$, $\overline{D}_\rho={\cal D}_\rho/D_0$ and $\overline{D}_s=D_s/D_0$. Finally, we assume  $\alpha=\frac{\rho }{\rho _{c}}-1 $ and $
\beta $ independent of $\rho$. The effect of
activity is assumed to affect the mean field phase transition only through the dependence
of the critical density $\rho _{c}$ on the magnitude of self-propulsion speed.
In this simplified form, the dynamics of the system is
characterized by two central parameters: the mean density  $\rho _{0}$ of active nematogens and
the  self-propulsion velocity $\overline{
v}=v_0/\sqrt{D_rD_0}$, which is effectively the Peclet number for this flow. The
other parameters $\gamma$, $\lambda$, $\overline{D}_Q$, $\overline{D}_\rho$ and $\overline{D}_s$ are in general functions
 of $\rho _{0}$ and $\overline{v}$, although we will treat them here as independent parameters and simply fix their values.

\section{Linear  Dynamics and Emergent Structures}
\label{sec:emergent}

The dynamics of self propelled rod-like particles with steric repulsion has been studied extensively
by numerical simulation of microscopic models~\cite{Mishra2006,Chate2006,Yang2010,Sam2012}.
This work has revealed a rich variety of emergent structures, including bands of high
density regions where the particles are ordered along the direction of the
band, lane formation, migrating defect structures and low Reynolds number
turbulence. Here we examine the minimal continuum model of  self-propelled nematic given by Eqs.~\eqref{4} to
identify the generic dynamical mechanisms responsible for the emergence of
these structures. As mentioned above, there are three important mechanisms for dynamical instabilities and parttern formation in these systems.
To unfold the role of each
of these mechanisms in controlling the large-scale dynamics of the system, we analyze the linear stability of the ordered nematic state  in various special cases that best highlight a particular mechanism.

The ordered nematic state has constant density $\rho_0$, zero mean polarization density, $\bm\tau_0=0$, and a finite value for the nematic alignment tensor.
Choosing a coordinate system with the $x$ axis pointing along the direction of broken nematic symmetry,
the alignment tensor in the uniform nematic state has components $Q^0_{xx}=-Q^0_{yy}=Q_0/2$ and $Q^0_{xy}=Q^0_{yx}=0$, with
$Q_0=\sqrt{\alpha_0/\beta}$ and $\alpha_0=\alpha(\rho_0)$. We now examine the linear stability of this state in various regions of parameters by considering the dynamics of small fluctuations,
$\delta\rho({\bf r},t)=\rho({\bf r},t)-\rho_0$, $\delta\bm\tau({\bf r},t)=\bm\tau({\bf r},t)$ and $\delta Q_{ij}({\bf r},t)=Q_{ij}({\bf r},t)-Q_{ij}^0$. We will generally work in Fourier space by introducing Fourier transforms of the fluctuations as
$\phi^\alpha_{\bf k}(t)=\int_{\bf r}e^{i{\bf k}\cdot{\bf r}}\delta\phi_\alpha({\bf r},t)$, where
$\delta\phi_\alpha=\left(\delta\rho,\bm\tau,\delta Q_{ij}\right)$.

\subsection{Dynamical Self-Regulation and Banding Instability}
\label{subsec:self-reg}

\begin{figure}[tbp]
\includegraphics [width=6cm] {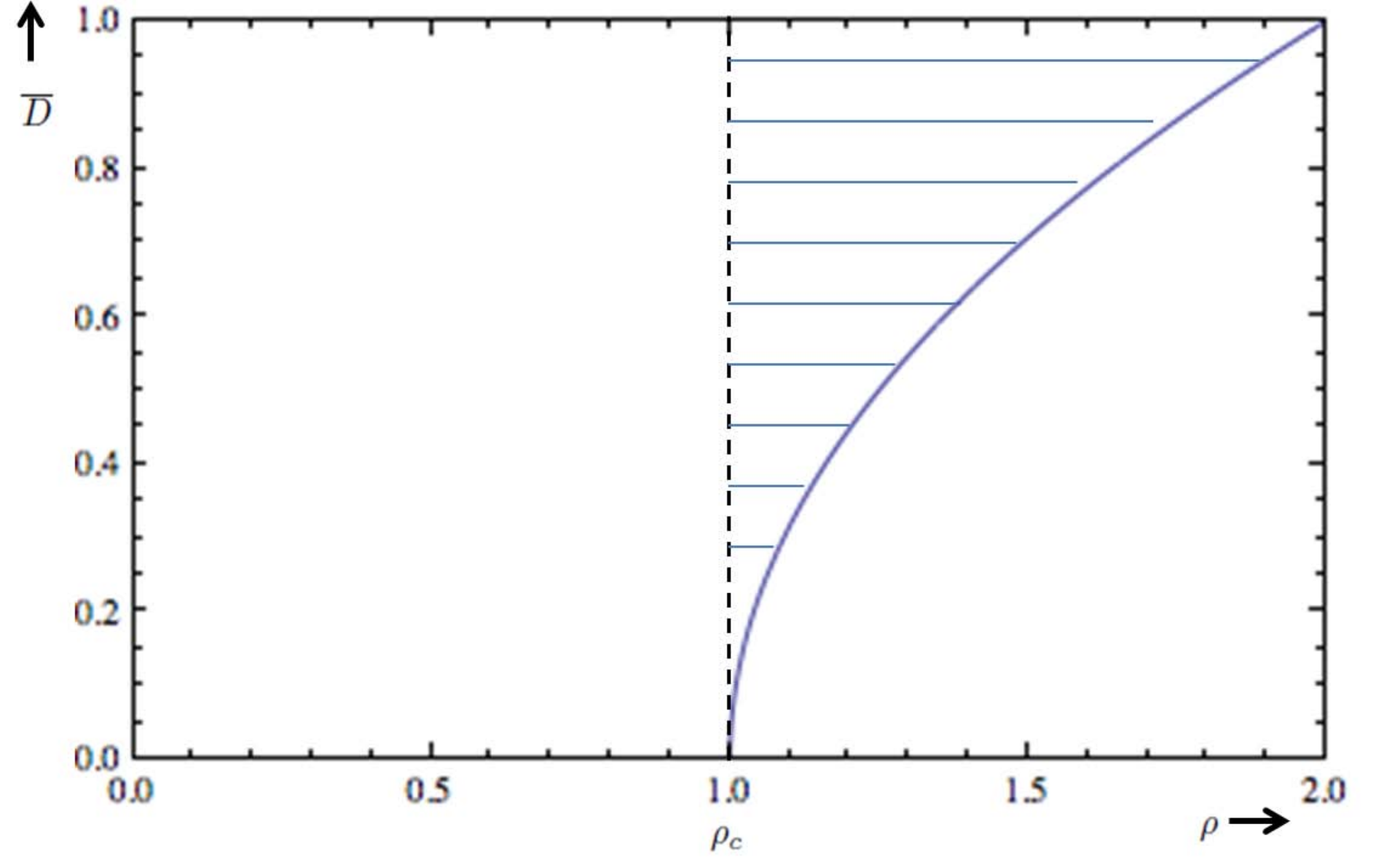}
\caption{ The banding instability that occurs due to the self-regulating nature of the flow. The striped region is the parameter space in which this instability occurs.}
\label{Fig3}
\end{figure}

We first consider the linear dynamics of the system in the region just above
the mean-field transition at $\rho_c$.  For simplicity we only discuss spatial variations normal
 to the direction of broken symmetry, as these correspond to the most unstable modes, i.e., let
$\mathbf{k}=k\mathbf{\hat{y}}$. Fluctuations in $\tau
_{x}$ and $\delta Q_{xy}$  then decouple and are always stable. The dynamics of
fluctuations in $\delta \rho $, $\tau _{y}$ and $\delta Q_{yy}$ is governed
by three coupled equations. Fluctuations in $\tau _{y}$
are always quickly damped near the mean-field transition, while the decay rate of $\delta Q_{yy}$, controlled to leading order by $\alpha _{0}$,
vanishes as $\rho _{0}\rightarrow \rho _{c}^{+}$. We therefore neglect fluctuations in $\tau _{y}$ and simply examine the
coupled dynamics of $\delta \rho $ and $\delta Q\equiv \delta Q_{yy}$, given
by
\begin{subequations}
\begin{gather}
\partial _{t}\delta \rho _{\mathbf{k}}=-k^{2}\delta \rho _{\mathbf{k}}-
\overline{D}_Qk^{2}\delta Q_{\mathbf{k}}  \label{6.1} \\
\partial _{t}\delta Q_{\mathbf{k}}=-\left[\frac{ \alpha _{0}}{2}+(1+\overline{D}
_{s})k^{2}\right] \delta Q_{\mathbf{k}}-\frac{1}{2}(\alpha ^{\prime
}Q_{0}+\overline{D}_\rho k^{2})\delta \rho _{\mathbf{k}}  \label{6.2}
\end{gather}%
where $\alpha ^{\prime }=\left( \frac{\partial \alpha }{\partial \rho }%
\right) _{\rho =\rho _{0}}$, or $\alpha ^{\prime }=1/\rho _{c}$ with the
chosen parameters. The decay of density and ordered parameter fluctuations
is then controlled by two coupled hydrodynamic modes. One of the modes has a
finite decay rate (proportional to $\alpha _{0}$) in the limit $k\rightarrow
0$ and is always stable. At small wavevector, the dispersion relation of the
other mode is given by
\end{subequations}
\begin{equation}
s_{y}(k)=-s_{2}k^{2}-s_{4}k^{4}+\mathcal{O}(k^{6})\;.  \label{7}
\end{equation}%
with $s_{2}=1-\frac{\overline{D}\alpha ^{\prime }}{\sqrt{\alpha _{0}\beta }}$
and $s_{4}>0$. Near the transition where $\alpha _{0}\rightarrow 0$, $s_{2}<0
$ and $s_{4}\simeq \frac{2\overline{D}^{2}\alpha ^{^{\prime }2}}{\alpha
_{0}^{2}\beta }$. As a result, $s_{y}(k)>0$ for a range of wavevectors,
resulting in the unstable growth of density and order parameter
fluctuations illustrated in Fig.~\ref{Fig3}. The fastest growing mode has wave vector $k_{0}=\sqrt{
-s_{2}/2s_{4}}\sim (\rho _{0}-\rho _{c})^{3/2}$. Including the coupling to $\tau_y$ will yield finite Peclet number corrections to the instability.
Note that this instability
is strongest in the vicinity of the order-disorder transition and is a
manifestation of the fact that the dynamics of the system is self
regulating, i.e., the control parameter associated with the phase
transition, namely the density is dynamically coupled to the emergent
ordering that results from the transition through the curvature induced
fluxes. This is the dynamics that leads the system to be intrinsically phase
separated \cite{Ramaswamy2003}.

We recall that polar active fluids exhibit a similar instability for wave
vectors parallel to the direction of mean order. In that case the mode that
goes unstable is a propagating mode and the instability signals the onset of
solitary waves consisting of alternating ordered and disordered bands
extending in the direction normal to that of mean order and traveling along
the direction of broken symmetry. These bands have been observed in
simulations of the Vicsek model~\cite{Gregoire2004,Chate2008}, as well as in
numerical solutions of the nonlinear hydrodynamic equations for polar fluids
~\cite{Bertin2006,Bertin2009,Mishra2010}. We have shown here that active nematics
exhibit a similar instability, controlled by the interplay of of curvature
currents ($\overline{D}_Q$) and the self-regulation due to the density
dependence of $\alpha $. The instability occurs even for $\overline{v}=0$,
i.e., is present in both active and self-propelled nematics. It occurs for
wavevectors perpendicular to the direction of broken nematic symmetry and
the mode that goes unstable is a diffusive one. It is therefore tempting to associate it with the emergence of the stationary bands
consisting of alternating ordered (nematic) and disorders regions that have
been seems in simulations of active systems with apolar interactions ~\cite
{Chate2006} and physical excluded volume interactions \cite{Yang2010},\cite{Sam2012}. Finally, but most importantly, this instability mechanism is \emph{generic}, in the sense that it does not depend
on microscopic parameters, but only on the presence of a dynamical feedback between density and active currents.

\subsection{Curvature Induced Flux}
\label{subsec:curvature}
Next we consider the region of small $\overline{v}$, $\lambda $
and $\gamma $, deep in the nematic phase. In this case, the long-wavelength
dynamics is controlled by hydrodynamic modes associated with fluctuations in
the density and the director $\mathbf{\hat{n}}$. This case has been
considered in the literature already and is summarized here for completeness
~\cite{Simha2002a,Ramaswamy2003,Baskaran2008}. For our choice of coordinates
to linear
order we have $\delta Q_{{xx}}=-\delta Q_{yy}=0$ and $\delta Q_{xy}=\delta
Q_{yx}=Q_{0}\delta \hat{n}\left( \mathbf{r},t\right) $. Neglecting
polarization fluctuations that decay on microscopic time scales, the
linearized equations are
given by
\begin{subequations}
\begin{gather}
\partial _{t}\delta \rho _{\mathbf{k}}=-k^{2}\delta \rho _{\mathbf{k}}-Q_{0}%
\overline{D}k^{2}\sin 2\theta \delta \hat{n}_{\mathbf{k}}\;,  \label{5.1} \\
\partial _{t}\delta \hat{n}_{\mathbf{k}}=-\frac{\overline{D}_\rho k^2}{2Q_{0}}\sin 2\theta
\delta \rho _{\mathbf{k}}+\left[ \overline{D}_{s}+\cos 2\theta \right]
k^{2}\delta \hat{n}_{\mathbf{k}}\;,  \label{5.2}
\end{gather}
\end{subequations}
where $\theta $ is the angle between $\mathbf{k}$ and the direction of broken symmetry ($x$). If $\theta=0,\pi$, the
two equations are decoupled and the modes are diffusive and stable.  For general $\theta$ one of the hydrodynamic modes becomes unstable for
$\overline{D}\overline{D}_\rho\sin^{2} 2\theta >2(\overline{D}_{s}+\cos 2\theta)$.
This can be satisfied provided  $\overline{D}_Q\overline{D}_\rho>2D_{s}$, i.e., the curvature driven
fluxes exceed the restoring effects of diffusion.
This instability has been discussed in detail elsewhere~\cite{Baskaran2008}.



\subsection{Induced Polar Order }
\label{subsec:polar}
Finally, we examine the effect of
fluctuations with spatial variations along the direction of broken symmetry,
i.e., $\mathbf{k}=k\mathbf{\hat{x}}$. The relevant coupled fluctuations in
this case are  $\delta \rho $, $\tau _{x}$ and $
\delta Q_{xx}$. For simplicity, we consider the regime of large Peclet
number $\overline{v}$, where the linear dynamics is controlled by Euler
order terms and neglect terms quadratic in the gradients, with the result
\begin{gather}
\partial _{t}\delta \rho _{\mathbf{k}}+ik\overline{v}\tau _{x,\mathbf{k}}=0
\label{7.1} \\
\partial _{t}\tau _{x,\mathbf{k}}+\gamma _{e}\tau _{x,\mathbf{k}}=-ik%
\overline{v}\delta Q_{xx,\mathbf{k}}-ik\frac{\overline{v}}{2}\delta \rho _{%
\mathbf{k}}  \label{7.2} \\
\partial _{t}\delta Q_{xx,\mathbf{k}}-\frac{\alpha _{0}}{2}\delta Q_{xx,%
\mathbf{k}}=ik\left( \overline{v}+\frac{\lambda }{2}Q_{0}\right) \tau _{x,%
\mathbf{k}}+\frac{\alpha ^{\prime }}{2}Q_{0}\delta \rho _{\mathbf{k}}
\label{7.3}
\end{gather}%
where $\gamma _{e}=1+\frac{\gamma }{2}Q_{0}^{2}-\frac{\gamma }{2}Q_{0}$.
As discussed earlier and highlighted in Fig.~\ref{Fig1.1}, the anisotropy of
small angle collisions in the nematic state enhances polar order by
suppressing the decay rate of $\tau _{x,\mathbf{k}}$ from its bare value of $
1$ (in units of $D_{r}^{-1}$) to $\gamma _{e}$. The dispersion relations
of the hydrodynamic modes associated with Eqs.~\eqref{7} are easily
calculated at small wavevectors. Clearly, if $\gamma _{e}\leq 0$,
In addition,
as a consequence of this built up of polar order, the
diffusive mode associated with the density fluctuations (i.e., the only
truly hydrodynamic mode in these considerations), given by
\begin{equation}
s_{x}(k)=-k^{2}\frac{\overline{v}^{2}}{2\gamma _{e}}\left( 1+\frac{%
2\alpha ^{\prime }}{\sqrt{\alpha _{0}\beta }}\right)   \label{8}
\end{equation}
becomes unstable. In general $\gamma _{e}$
depends on microscopic details of the model, but there is no reason to
exclude a priori that it could change sign and indeed does for the case of
long thin hard rods with excluded volume interactions \cite{Baskaran2008}.
The linear analysis here is of limited utility because of the existence of a
homogeneous instability but is shown here to indicate that the build up of
polarization due to the momentum conserving nature of the interactions has a dramatic consequence on the dynamics of the system.
This may indeed be the mechanism responsible for the smectic order observed within a single polar cluster in simulations
 of self-propelled rods~\cite{Yang2010,
Sam2012}. Finally, we stress that the nonlinear homogeneous terms
proportional to $\gamma $ and responsible for the renormalization of $\gamma
_{e}$ always vanish in an equilibrium state because the nematic symmetry
of such a state by definition forbids a nonzero uniform value of the mean
polarization.
%

\section{Discussion}
\label{sec:discussion}

We have considered in this paper the hydrodynamics of active overdamped fluids that can order in nematic states.
These are collections of active particles that interact via apolar (nematic) aligning interactions, such as steric repulsion or medium-mediated hydrodynamic couplings.
One can identify two classes  of such fluids, depending on the properties of the individual active units. Active nematics consist of shaker particles that are themselves apolar. Self-propelled nematics are collections of  particles that are physically head-tail symmetric (such as SP rods), but where a microscopic dynamical polarity is induced by self-propulsion . Although both systems form ordered states of nematic symmetry, their dynamical behavior is qualitatively different, as seen in recent simulations~\cite{Peruani2006,Ginelli2010,Yang2010}.

The hydrodynamic equations of active nematics have the form given in Eqs.~{2}. We have shown that the same equations are also obtained by considering SP particles and neglecting the effect of self-propulsion on the interaction between active units, suggesting that the active nematic may be considered the zero Peclet number $\overline{v}$ limit of self-propelled nematic. In this case the only active term is the curvature current proportional to ${\cal D}_Q$ in Eq.~\eqref{2.1}. This non equilibrium coupling of orientation and flow induces instabilities of the ordered state that have been studied before in the literature~\cite{Simha2002,Ramaswamy2003,Baskaran2008} and are also summarized in section. \ref{subsec:curvature}. The curvature current is also key in controlling the banding instability arising from dynamical self-regulation discussed in section~\ref{subsec:self-reg}. In fact this instability, although not discussed before in the literature for overdamped active nematic, occurs in all active fluids of nematic symmetry, both for shakers and self-propelled particles. It arises from the density dependence of the parameter $\alpha(r\rho)$ that controls the mean-field transition and the fact that in active systems $\rho$ is not tuned from the outside, as in equilibrium, but is itself a dynamical variable convected by the order parameter.

The hydrodynamic equations of self-propelled nematics  given in Eqs.~\eqref{3} (or Eqs.~\eqref{4} in the dimensionless form studied here) contain many new active terms that arise from modifications of the two-body interaction due to self propulsion. These equations have also been derived by us for a specific microscopic model of self-propelled hard rods~\cite{Baskaran2008a,Baskaran2010}, although the low order closure of the kinetic theory used in that work only gives terms up to quadratic in the hydrodynamic fields. Self-propelled nematics also exhibit both the curvature induced instability discussed in section \ref{subsec:curvature} and the banding instability discussed in section \ref{subsec:self-reg}. Both are of course modified at finite Peclet number due to additional convective contributions to the underlying mechanisms the details of which will be discussed elsewhere. In addition, self-propulsion yields a novel instability due to the built-up of local polar order discussed in section \ref{subsec:polar}. This  arises because in the nematic state most binary collisions involve nematogens that are nearly aligned or anti aligned, as shown in Fig.~\ref{Fig1.1}. When the nematogens are self-propelled, collisions of nearly aligned and nearly anti aligned pairs are not identical. Nearly aligned pairs tend to further align upon collisions, while nearly anti-aligned pairs are turned away from each other. As a result, local polar order is enhanced and the nematic state becomes unstable as discussed in section~\ref{subsec:polar}. It is tempting to associate this instability with the onset of ``polar clusters" that have been observed ubiquitously in simulations of self-propelled rods~\cite{Peruani2006,Ginelli2010,Yang2010}, as well as in experiments in gliding myxobacteria~\cite{Peruani2012}.

\begin{acknowledgments}
MCM was supported by the National Science Foundation through awards DMR-0806511 and DMR-1004789. AB was supported by the
Brandeis-MRSEC through NSF DMR-0820492.
\end{acknowledgments}

\bibliography{Baskaran-MCM-EPJE-2012}

\end{document}